# Landau Diamagnetism Revisited

**Sushanta Dattagupta[†,**], Arun M. Jayannavar[‡] and Narendra Kumar[#]**

[†]S.N. Bose National Centre for Basic Sciences, Block JD, Sector III, Salt Lake, Kolkata 700 098, India
[‡]Institute of Physics, Sachivalaya Marg, Bhubaneswar 751 005, India
[#]Raman Research Institute, Bangalore 560 080, India

The problem of diamagnetism, solved by Landau, continues to pose fascinating issues which have relevance even today. These issues relate to inherent quantum nature of the problem, the role of boundary and dissipation, the meaning of thermodynamic limits, and above all, the quantum–classical crossover occasioned by environment-induced decoherence. The Landau diamagnetism provides a unique paradigm for discussing these issues, the significance of which is far-reaching. Our central result connects the mean orbital magnetic moment, a thermodynamic property, with the electrical resistivity, which characterizes transport properties of material.

In this communication, we wish to draw the attention of the reader to certain enigmatic issues concerning diamagnetism. Indeed, diamagnetism can be used as a prototype phenomenon to illustrate the essential role of quantum mechanics, surface–boundary, dissipation and nonequilibrium statistical mechanics itself.

Diamagnetism is a material property that characterizes the response of an ensemble of charged particles (more specifically, electrons) to an applied magnetic field. The magnetic field causes cyclotron motion of each particle, thereby creating an orbital magnetic moment, governed by Faraday–Lenz's law. Thus the system exhibits a negative magnetic susceptibility, the hallmark of diamagnetism. What is however remarkable is that diamagnetism is calculated to be zero within the framework of classical Gibbsian statistical mechanics. This result goes under the celebrated Bohr–Van Leeuwen's (BV) theorem[1]. What is even more remarkable is that the boundary of the enclosure, as also any internal boundary, plays a crucial role; the bulk contribution of the diamagnetic moment exactly cancels the contribution arising from the so-called 'skipping orbits' of those electrons which hit the boundary and get multiply reflected to constitute what is called an 'edge current'[2].

Thus it was a great triumph of quantum mechanics when Landau showed in 1930 that the discreteness of energy levels and the consequent degeneracy of each level led in a natural way to diamagnetic susceptibility[3]. Landau's calculation also suggests that in quantum mechanics the bulk and the surface contributions are different, though opposite in sign, and therefore, the cancellation of the two terms is incomplete, unlike in the classical case. This fascinating result which brings the boundary of a container, normally passive in determining thermodynamic properties, to essential reckoning, led Peierls to term diamagnetism as one of the surprises in theoretical physics[4]. Landau's pathbreaking result also demonstrated that the calculation of diamagnetic susceptibility did indeed require an explicit quantum treatment.

Turning to the classical domain, two of the present authors had worried, some years ago, about the issue: does the BV theorem survive dissipation[5]? This is a natural question to ask as dissipation is a ubiquitous property of materials, and in the present context, can be viewed to occur as a result of inelastic scattering. Now, dissipation or damping necessarily requires a time-dependent analysis. Naturally, therefore, the calculation was set in the framework of the Langevin/Fokker–Planck equations, just as in the case of Brownian motion, with the Lorentz-force appearing as a systematic drift. Because the calculational approach had built-in, at the outset, the fluctuation–dissipation theorem, the diamagnetic moment was extracted as a stationary, (i.e. asymptotic, time going to infinity) property. Interestingly, surprise was in store, as the diamagnetic moment was found to be non-zero and dependent on the coefficient of friction, seemingly at odds with the BV theorem. This result however, turned out to be one red herring, as it were, when the boundary was treated carefully. Thus we discovered the curious result: when the stationary limit was taken first and the thermodynamic (i.e. volume going to infinity) limit taken next, the BV theorem was restored, while it was violated when the two limits were interchanged – again bringing in the issue of the boundary in an explicit manner.

We stated earlier that diamagnetism is an essential quantum mechanical property. Thus it was natural to extend the Jayannavar–Kumar analysis to the quantum domain[6]. The resultant treatment, based on what is being branded now as dissipative quantum mechanics, again brings to the fore certain remarkable phenomena which we wish to focus onto. Four distinct issues were tackled in our earlier work[6]: (a) approach to equilibrium of a quantum dissipative system, the analysis of which brings out the subtle role of boundary electrons, a key point, as mentioned before, in the whole business of orbital magnetism; (b) the effect of dissipation on Landau diamagnetism, an equilibrium property; (c) quantum–classical crossover as the system transits from the Landau to the Bohr–Van Leeuwen regimes as a function of damping; and (d) the combined effect of dissipation and confinement on Landau diamagnetism, the latter arising from coherent cyclotron motion of the electrons. The item (d) is particularly relevant in the context of intrinsic decoherence in mesoscopic structures in view of heat bath-induced influence[7]. Further, items (b)–(d), put together, had prompted us to ask question on not whether the BV theorem survives dissipation, but whether the Landau diamagnetism itself survives dissipation?

---

**For correspondence. (e-mail: sdgupta@boson.bose.res.in)

The central result for the diamagnetic moment $M$ per particle of a non-degenerate gas of free electrons, in contact with an Ohmic bath, derived in Dattagupta and Singh[6] can be written as

$$M = -\frac{2KT}{B}\omega_c^2 \sum_{n=1}^{\infty} \frac{1}{(v_n + r)^2 + \omega_c^2}, \quad (1)$$

where

$$v_n = \frac{2\pi}{\hbar} KTn, \quad (2)$$

and

$$\omega_c = \frac{eB}{mc}, \quad (3)$$

$B$ being the magnetic field. We would like to remind the reader, in the context of the point (a) in the paragraph above, that the result, eq. (1), was extracted from the solution of the appropriate quantum Langevin equations, by taking the $t \to \infty$ limit first and then letting the boundary of the system go to infinity. The complete expressions, including the time-dependence and the dependence on the parameters of the confining boundary, from which eq. (1) has been deduced, are available in the literature[6]. It is also evident that in the limit of zero damping ($r = 0$), eq. (1) reduces to the Landau answer:

$$M^o = \frac{e\hbar}{2mc}\left[\frac{2KT}{\hbar\omega_c} - \coth\left(\frac{\hbar\omega_c}{2KT}\right)\right]. \quad (4)$$

It may be recalled that eq. (4) is usually derived from the partition function of a gas of free electrons in the presence of a magnetic field[2], whereas eq. (1) is extracted from a non-equilibrium method[6]. The calculation itself demonstrates *inter alia* that the path to equilibrium from a non-equilibrium state is not unique – it is important as to whether the system is 'allowed to equilibrate' in the presence of a boundary or in free space!

Another significant point concerning eq. (1), which provided one of the motivations for us to write this note, is the explicit presence of the friction coefficient $r$ in an equilibrium function $M$. Recall that $r$ had its origin in the coupling constant, characterizing the interaction between the gas of electrons and the bath. Normally, such a constant disappears from equilibrium answers, which are extracted from irreversible statistical mechanical methods, wherein the system–bath interaction is treated as a 'weak coupling'. What is crucial for dissipative diamagnetism is that the system–bath interaction has to be treated exactly – there is no clear-cut separation between what is a system and what is a bath – both are inexorably linked into one many-body system!

A related point is that the same coefficient $r$ determines what is however a transport property, viz. the Drude resistivity:

$$v = \frac{mr}{\rho e^2}, \quad (5)$$

$\rho$ being the change carrier number density. Eliminating $r$ in favour or $v$ and introducing the 'Hall resistivity' $R$ as

$$R = \frac{B}{\rho ec}, \quad (6)$$

eq. (1) can be suggestively rewritten as

$$-\frac{mc}{e\hbar}M = \frac{1}{v}\sum_{n=1}^{\infty}\frac{1}{1+(\bar{\mu}_n + \bar{v})^2}, \quad (7)$$

where

$$v = \frac{\hbar\omega_c}{2KT}, \quad (8)$$

$$\bar{v} = \frac{v}{R},$$

and

$$\bar{\mu}_n = \frac{\pi n}{v}. \quad (9)$$

Equation (7) is a novel result: transport characterized by the resistivity $\bar{v}$ determining an equilibrium property of orbital moment $M$ in an efficient manner, because resistance, in some sense, incorporates all relevant material properties. From another point of view, and especially in the context of contemporary relevant mesoscopic structures, larger the resistance, higher is the level of incoherence. Thus, while Landau diamagnetism in eq. (4) is a result of coherent property of a quantum system, dissipative diamagnetism, captured by eq. (7), is an expression of intrinsic incoherence in a macroscopic quantum system. To illustrate this point, we plot in Figure 1, the magnetic moment versus the resistivity: the increase in resistivity with the concomitant enhancement of incoherence leads to a vanishing magnitude of the orbital moment, as though Bohr and Van Leeuwen are resurrected!

We may remark in passing that the Drude formula for resistivity, eq. (5), is robust, almost oblivious to whether $r$ originates from scattering with phonons or defects or other electrons[8]. Thus eq. (5) remains valid in general, except that the interpretation and calculation of $r$ become an increasingly complex task, when the effects of disorder and electron correlations are to be considered. Therefore, our relation between a thermal equilibrium property, viz. $M$ and a transport property, viz. $\bar{v}$, though appears facile, as it is based on perhaps the simplest of models, subsumes a deep connection between orbital magnetism and dissipation – it transcends detailed issues of disorder and concomitant localization effects, as well as strong Coulomb correlations between electrons.

Our final remark is in the context of quantum–classical crossover due to environment-induced decoherence, epitomized by the expression given in eq. (7). Remember that there are two distinct quantum effects which, in the model of an electron gas, can be described thus: (1) one is due to the fact that the position and momentum of an electron cannot be determined simultaneously, which is why a continuum of states splits into discrete Landau levels; we call this the phase space factor, and (2) the other is due to the quantum statistics of indistinguishable particles, in this case the Fermi–Dirac one. As long as the de Broglie wavelength is smaller than the average inter-electron distance, the effect of statistics or the Pauli exclusion principle can be neglected, which is indeed what was done in deriving eq. (4) for a non-degenerate gas. Now, in examining the issue of quantum–classical crossover, it is interesting to introspect on what gets classicalized by environment-induced decoherence first: the phase space factor or the quantum statistics? Inasmuch as, in the calculation enumerated above, the electron gas was treated as non-degenerate, i.e. the thermal de Broglie wavelength of the electron was taken to be much less than the mean electron–electron spacing, the quantum statistics was already rendered ineffective (or classical). It was, therefore, only the phase space factor that got obfuscated by the increasing decoherence effects. On the other hand, it would be extremely interesting to look for another quantum phenomenon, which is intrinsically due to statistics, such as the Bose–Einstein condensation, to enquire what dissipation does to it! It may well be that it is the phase-breaking length rather than the de Broglie length that must exceed the inter-particle spacing as a precondition for the quantum statistics to be effective.

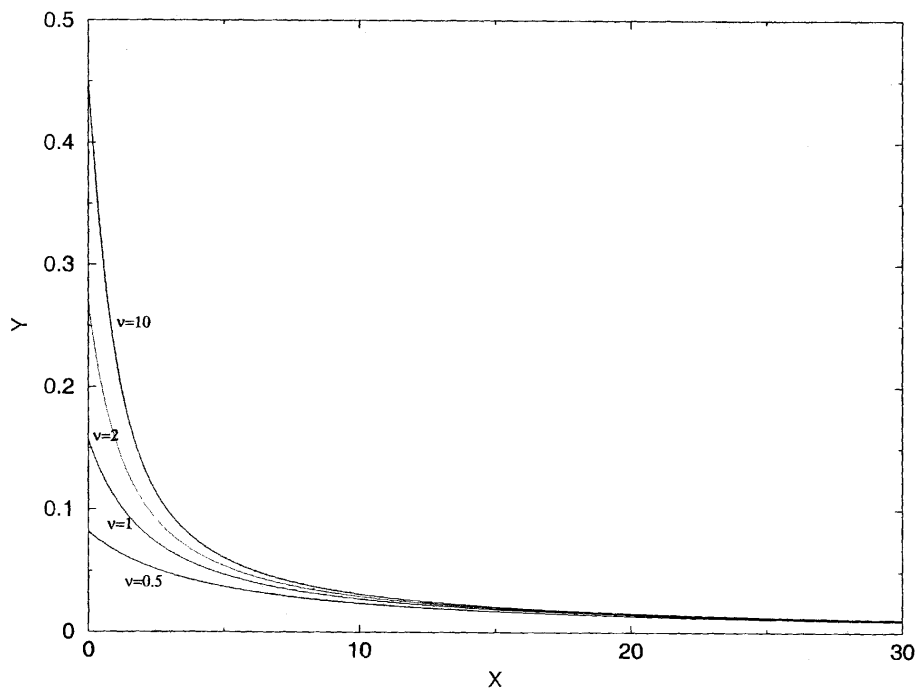

**Figure 1.** Scaled diamagnetic moment $Y = -(mc/e\hbar)M$ plotted against scaled resistivity $X = \bar{v}$, for different values of the scaled cyclotron frequency $v$.

Summarizing, diamagnetism seems to be a unique material property which does not exist in classical mechanics, but lives only in the quantum domain. Since quantum mechanics has to do with the phase of the wavefunction, diamagnetism can be viewed to arise from quantum coherence. On the other hand, all materials are inherently dissipative due to scattering of electrons off phonons, impurities, other electrons, etc. Because dissipation leads to incoherence, making a quantal system look seemingly classical, it is interesting to study the influence of dissipation on diamagnetism. Such an analysis, summa-rized in this article, elucidates many subtle issues of contemporary condensed matter physics.

ACKNOWLEDGEMENT. We thank Kamal Saha for help in generating Figure 1.